\documentclass[prl, amsmath, amssymb, amsfonts, reprint, groupedaddress, aps]{revtex4-2}
\usepackage{graphicx}
\usepackage{float}
\usepackage{subfloat} 
\usepackage{caption}
\usepackage{subcaption}
\usepackage{dcolumn}
\usepackage{bm}
\usepackage[colorlinks=true,linkcolor=blue]{hyperref}
\usepackage{url}
\usepackage{amsmath}
\allowdisplaybreaks
\begin{document}

\title{Entangled laser beams and quantum ghost frequency comb}
\author{Yanhua Shih}
\email{shih@umbc.edu}
\affiliation{Department of Physics, University of Maryland Baltimore County, Baltimore, Maryland 21250, USA}

\begin{abstract}
This letter reports on the study of entangled laser beams, or entangled coherent states, from their generation to their nonlocal coherent behavior.  Although in continuous wave operation, the entangled laser beams are able to produce comb-like correlation with 100\% contrast in distant joint photodetection. We name this comb-function quantum ghost frequency comb (QGFC). What is the cause of these periodic sharp correlations?  Can we trust zero-coincidences, or anti-correlation, in the joint measurement of CW laser beams?  Besides its fundamental interests, bright QGFCs make important contributions to the fields of nonlocal precision spectroscopy, positioning, and time transfer. Superior to entangled photon pairs, measurements of entangled laser beams do not rely on photon counting and can be performed over greater distance in shorter time with higher resolution and accuracy.
\end{abstract}

\maketitle

Since Einstein-Podolsky-Rosen (EPR) suggested a \emph{gedankenexperiment} and introduced an entangled two-particle system in 1935 \cite{EPR}, we have experimentally verified the nonlocal EPR correlation of entangled photon pairs created in positronium annihilation \cite{Wu}, in atomic cascade decay \cite{Clauser-1, Clauser-2, Fry, Aspect-1, Aspect-2}, and in spontaneous parametric down-conversion (SPDC) \cite{Alley-Shih, Kwiat}.  The state of the signal-idler photon pair created from SPDC is a typical EPR state, 
\begin{align}\label{state-3}
\left| \Psi \right\rangle &= \int d {\bf k}_{s} \, d {\bf k}_{i} \, \delta \left( \omega
_{s}+\omega _{i}-\omega _{p}\right) \delta \left( {\bf k}_{s}+{\bf
k}_{i}-{\bf k} _{p}\right) \cr
& \hspace{20mm}\times a_{s}^{\dagger }({\bf k}_{s})\, a_{i}^{\dagger }({\bf k}_{i}) \, | 0\rangle 
\end{align}
where $a_{s}^{\dagger }({\bf k}_{s})$ and $a_{i}^{\dagger }({\bf k}_{i})$ are the creation operators, 
$\omega_{j}$ and ${\bf k}_{j}$, $j = s, i, p$, are the
frequency and wavevector of the signal ($s$), idler ($i$), and the pump ($p$). In Eq.~(\ref{state-3}), we assumed
monochromatic single-mode pump. Loosely speaking, in the nonlinear interaction, two photons, namely the signal and idler, are created in pairs with conserved energy and momentum.  The most interesting and astonishing behavior of the signal photon and the idler photon is their nonlocal second-order coherence or correlation observable from their joint detection. For example, the signal photon and the idler photon can be released from the source at arbitrary angles to the left and right, but if the signal photon is observed at a space-time point on the left, the idler photon must be found at a unique spacetime point on the right, no matter how far apart the two are. Mathematically, this nonlocal correlation is represented by a 100\% contrast $\delta$-function like function of the space-time interval between the two photodetection events. Bell proved that this kind of $>$71\% contrast correlation represents a nonlocal quantum behavior that cannot be explained by any classical statistical theory \cite{Bell}. In fact, before the observation of EPR correlation, Hanbury Brown and Twiss (HBT) had discovered the ``classical" correlation of thermal light, a 50\% contrast $\delta$-function like function of the space-time interval between two distant photodetection events \cite{HBT}.   

Recent experimental observation of ghost frequency comb (GFC) of CW laser beams has attracted much attention from both fundamental and practical perspectives \cite{GFC1, GFC2}. Unlike a conventional frequency comb, in that experiment, the CW laser beam does not consist of a pulse train but instead it is in a continuous-wave operation. The laser beam was split into two, each sent to a distant observer. In their local measurements, both observers observe constant intensity with no pulse structure present. Surprisingly, a 50\% contrast pulse train of comb-like, ultra-narrow peaks is observed from the joint measurement of the distant CW laser beams \cite{GFC1, GFC2}. The laser beam, consisting of half a million longitudinal modes, is created from a $\sim$10 kilometer fiber ring cavity. The state of each mode can be approximated as a coherent state, represents a group of indistinguishable photons. Half a million coherent states are randomly distributed in the phase space.  Although this comb-function was named GFC, similar to the historical HBT correlation, its 50\% contrast may not be considered a quantum correlation in Bell's theory \cite{Bell}. In this letter we ask a question: Is it possible to create and entangle two laser beams and observe quantum correlation with 100\% contrast? 

\begin{figure}[htb]
    \centering
    \includegraphics[width = 1.05 \linewidth]{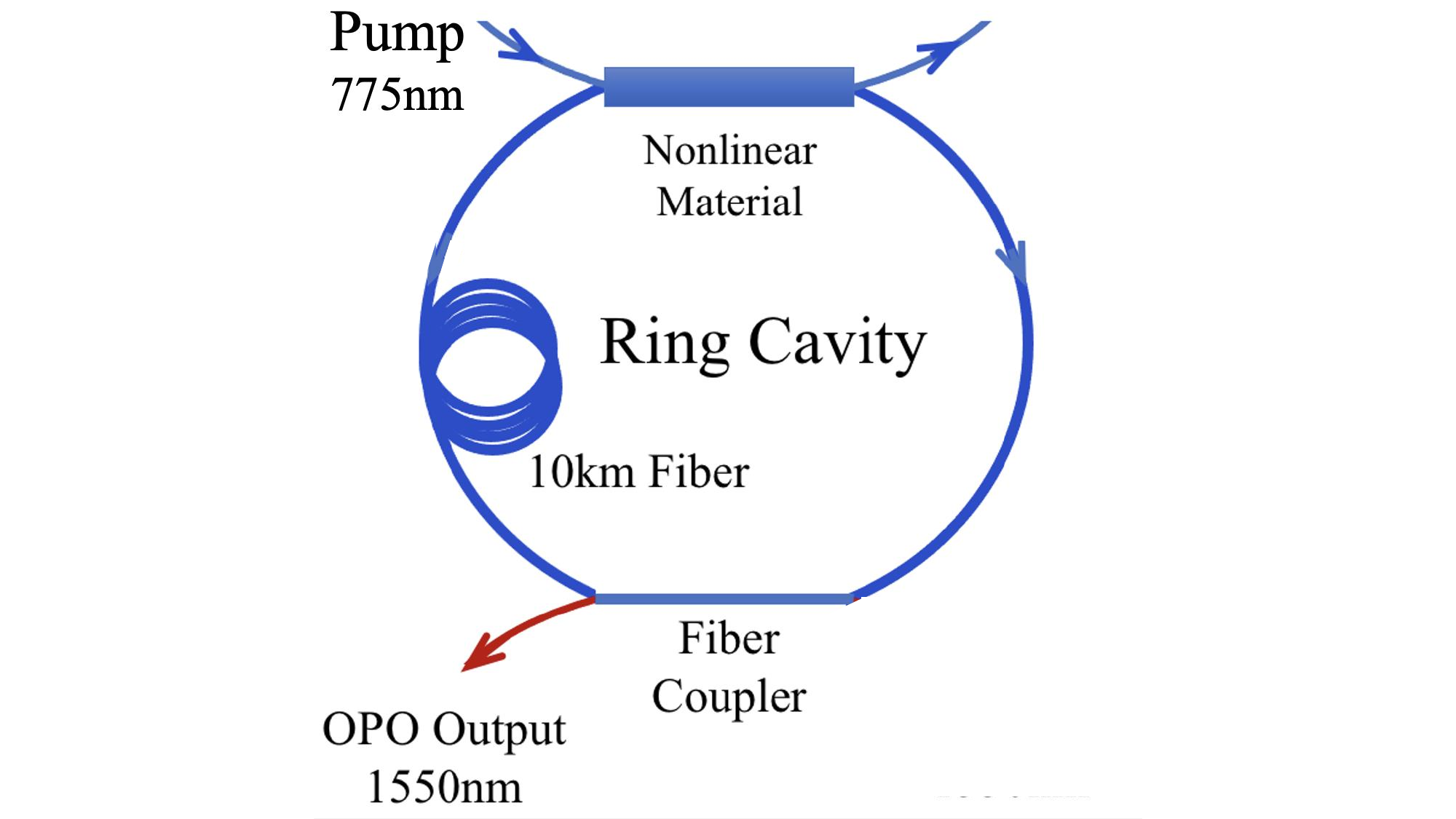}
    \caption{Schematic of a fiber Optical Parametric Oscillator (OPO). A $\sim$10 km fiber cavity is able to produce 
    $\sim$500,000 entangled pairs of longitudenal-cavity-modes within $\sim$10 GHz bandwidth of phase matching. 
    Half-million pairs of signal-idler modes added in an entangled and coherent manner form the state of Eq.~(\ref{state-2}).
    \label{fig:OPO-10km}}
\end{figure}

In light of new fiber technologies, we conclude that it is possible to generate a large number of entangled longitudinal-cavity-modes from an optical parametric oscillator (OPO) such as that shown in Fig.~\ref{fig:OPO-10km}. The coherent summation of these mode-pairs produces a state similar to Eq.~(\ref{state-3}) 
\begin{align}\label{state-2}
\left| \Psi \right\rangle &= \prod_{s,i}\delta \left( \omega_{s}+\omega _{i} - \omega _{p}\right) 
\delta \left( {\bf k}_{s}+{\bf k}_{i}-{\bf k} _{p}\right) \cr
& \hspace{10mm}\times |\alpha_{s}({\bf k}_{s})\rangle 
|\alpha_{i}({\bf k}_{i})\rangle
\end{align}
where $|\alpha_{s}({\bf k}_{s}) \rangle$ and $| \alpha_{i}({\bf k}_{i}) \rangle$, $|\alpha| \gg 1$, are the state vectors 
of the signal-idler of the OPO in the coherent state representation \cite{Glauber}; and $\omega_{j}$ and ${\bf k}_{j}$, for $j = s, i, p$, are the frequency and wavevector of the signal ($s$), idler ($i$), and the monochromatic single-mode pump ($p$) laser beam.  Considering the signal and the idler, respectively, a group of indistinguishable photons in the same coherent state, the state of Eq.~(\ref{state-2}) indicates: (1) each group of indistinguishable photons, i.e., a signal represented by $| \alpha_{s}({\bf k}_{s}) \rangle$, is entangled with another unique group of indistinguishable photons, i.e., an idler represented by $|\alpha_{i}({\bf k}_{i} \rangle$, by means of energy conservation $\hbar\omega_{s}+ \hbar\omega _{i} = \hbar\omega _{p}$ and momentum conservation $\hbar{\bf k}_{s}+\hbar{\bf k}_{i} =\hbar{\bf k} _{p}$; and (2) a large number of such signal-idler pairs are added constructively, and phase matched with the monochromatic single-mode pump, in a coherent manner to form a pair of entangled signal-idler laser beams. 

The study of Optical Parametric Amplifiers (OPAs) and OPOs have been used with great success to generate squeezed states \cite[and the references therein]{OPO-Rev}. Squeezing effects have also been studied intensively and successfully in terms of squeezed continuous variables \cite{Squeezed-000, Squeezed-00, Squeezed-01, Squeezed-02, Squeezed-03, Squeezed-04, Squeezed-05}. Squeezing from OPO operating above threshold has been demonstrated recently as well \cite{Squeezed-1, Squeezed-2, Squeezed-3, Squeezed-4}. It is not difficult to find from Eq.(\ref{state-2}) the physics of entangled coherent states or entangled laser beams, generated from an OPO with millions of cavity-mode-pairs, is very different from that of the two-mode squeezed state.  A large number of entangled mode-pairs is crucial for generating the periodic and sharp correlation peaks of QGFC. A bright entangled light source is of great value in both fundamental and practical research. Superior to entangled photon pairs, measurements of entangled laser beams do not rely on photon counting and can be performed over much greater distance in shorter time with higher resolution and accuracy.
By utilizing entangled laser beams, we expect to achieve more in fundamental research and practical applications. 

The state of the signal-idler pair of an OPO can be calculated, quantum mechanically, 
by the perturbation theory with the help of the nonlinear interaction
Hamiltonian.  The nonlinear interaction arises in a nonlinear material driven by a
pump laser beam.  The polarization, i.e., the dipole moment per
unit volume, is given by
\begin{align}
P_{i}=\chi^{(1)}_{i,j}E_{j}+\chi^{(2)}_{i,j,k}E_{j}E_{k}+
\chi^{(3)}_{i,j,k,l}E_{j}E_{k}E_{l}+...
\end{align}
where $\chi^{(m)}$ is the $m$th order electrical susceptibility
tensor.  In the case of bi-mode state, it is the second order nonlinear susceptibility
$\chi^{(2)}$ that plays the role.  The second-order nonlinear
interaction Hamiltonian can be written as
\begin{equation}
H=\epsilon_{0}\int_{V}d\mathbf{r}\ \chi^{(2)}_{ijk}\ E_{i}E_{j}E_{k}
    \label{H}
\end{equation}
where the integral is taken over the interaction volume, $V$.

It is convenient to use the Fourier representation for the
electrical fields in Eq.~(\ref{H}):
\begin{align}\label{fourier}
{\bf E}({\bf r},\,t)=\int d{\bf k}\ {\bf E}({\bf k})
e^{-i(\omega({\bf k})t-{\bf k}\cdot{\bf r})} + h.c.
\end{align}
Substituting Eq.~(\ref{fourier}) into Eq.~(\ref{H}) and keeping only the
terms of interest, we obtain the three-wave mixing Hamiltonian in the
interaction representation:
\begin{align}\label{HH}
\hat{H}_{\textrm{int}}(t)=\epsilon_{0}\int_{V}d \mathbf{r}\int d {\bf
k}_{s}\,d {\bf k}_{i} \, \chi_{l m n}^{(2)}
\hat{E}_{p l}^{(+)}e^{i(\omega_{p}t-{\bf k}_{p}\cdot{\bf r})} \nonumber \\
\hat{E}_{s m}^{(-)}e^{-i(\omega_{s}({\bf k}_{s})t-{\bf k}_{s}\cdot{\bf r})}
\hat{E}_{i n}^{(-)}e^{-i(\omega_{i}({\bf k}_{i})t-{\bf k}_{i}\cdot{\bf
r})}+h.c.,
\end{align}
to simplify the calculation, we have also assumed the pump field to be plane and
monochromatic with single wave vector ${\bf k}_{p}$ and frequency
$\omega_{p}$.

It is easily noticeable that the volume
integration in Eq.~(\ref{HH}) can be done for some simplified cases.  At this point,
to simplify the calculation, we assume that $V$ is infinitely large.   
The interaction Hamiltonian in Eq.~(\ref{HH}) is written as
\begin{align}\label{Hi1}
& \hat{H}_{\textrm{int}}(t) =\epsilon_{0}\int d {\bf k}_{s}\,d {\bf k}_{i}\,
\chi^{(2)}_{lmn}\, \hat{E}_{p\,l}^{(+)} \hat{E}_{s\,m}^{(-)}\hat{E}_{i\,n}^{(-)} \cr
& \ \ \times \delta({\bf k}_{p}-{\bf k}_{s}-{\bf k}_{i})
e^{i(\omega_{p}-\omega_{s}({\bf k}_{s})-\omega_{i}({\bf
k}_{i}))t}+h.c. 
\end{align}
It is reasonable to consider the pump field a classical wave (a laser beam), and quantize the signal and idler fields,
\begin{align*}
\hat{E}^{(-)}({\bf k}) = i\sqrt{\frac{2\pi\hbar\omega}{V}}\hat{a}^{\dagger}({\bf k}), \ 
\hat{E}^{(+)}({\bf k}) = -i\sqrt{\frac{2\pi\hbar\omega}{V}}\hat{a}({\bf k})
\end{align*}
where $a^{\dagger}({\bf k})$ [$\hat{a}({\bf k})$] is photon creation [annihilation]
operator. The state of the signal-idler mode pair
can be calculated by applying the standard quantum theory of perturbation \cite{ShihBook}. 

In the following, we simplify the state calculation into three steps as an approximation: 

(1) Consider the first nonlinear interaction, i.e., SPDC, creates a set of entangled mode pairs simultaneously,
\begin{align*}
\big[ |0\rangle +  a_{s}^{\dagger }(\omega_{s})\, a_{i}^{\dagger }(\omega_{i}) \, | 0\rangle \big]_
{\omega_s + \omega_i = \omega_p}
\end{align*}    
where we assumed single transverse and longitudinal modes of signal, $\omega_s$, and idler, $\omega_i$. 
Not to introduce any confusion, the integrals are removed, and mark the phase matching condition with each mode pair as a parameter index.  In the case of SPDC, since all signal-idler mode pairs must be phase matched with the monochromatic single mode pump, all created mode pairs of SPDC added constructively in an entangled and coherent manner form the state 
\begin{align}\label{state-2-10}
\left| \Psi \right\rangle = \prod_{s,i} \big[ |0\rangle +  a_{s}^{\dagger }(\omega_{s})\, a_{i}^{\dagger }(\omega_{i}) \, | 0\rangle \big]_{\omega_s + \omega_i = \omega_p}.
\end{align}
Taking the first-order approximation, we thus have
\begin{align}
\left| \Psi \right\rangle = \sum_{s,i} \delta \left( \omega
_{s}+\omega _{i}-\omega _{p}\right) a_{s}^{\dagger }(\omega_{s})\, a_{i}^{\dagger }(\omega_{i}) \, | 0\rangle 
\end{align}
which is the 1-D version of Eq.~(\ref{state-3}).

(2) Consider the second, third, ... $n$th nonlinear interactions, in the case of OPO, the \emph{stimulated} parametric amplification add the \emph{stimulatory} creations constructively.  The state of each cavity-mode can be 
approximated as 
\begin{align}\label{state-2-2}
& \ \ \ \ |\Psi\rangle \simeq \big[ |0\rangle +  
 a_{s}^{\dagger }(\omega_{s})\, a_{i}^{\dagger }(\omega_{i}) \, | 0\rangle \big]^n_
{\omega_s + \omega_i = \omega_p} \cr
& \simeq |0\rangle +\frac{n}{1!} \, a_{s}^{\dagger }(\omega_{s})\, a_{i}^{\dagger }(\omega_{i})  |0 \rangle  \cr
& \ + \frac{n(n-1)}{2!} \, [a_{s}^{\dagger }(\omega_{s})\, a_{i}^{\dagger }(\omega_{i}) )]^2 |0\rangle \cr
& \  + \frac{n(n-1)(n-2) ...(n-m+1)}{m!} \, [a_{s}^{\dagger }(\omega_{s})\, a_{i}^{\dagger }(\omega_{i}) ]^m  |0 \rangle \cr
& \ + ... \cr
& \simeq \big[ |\alpha_s(\omega_s) \rangle | \alpha_i(\omega_i) \rangle \big]_{\omega_s + \omega_i = \omega_p}. 
\end{align}
(3) Since all signal-idler mode pairs must be phase matched with the monochromatic single mode pump, all the
\emph{stimulatory created} mode pairs added constructively in an entangled and coherent manner form the state 
of Eq.~(\ref{state-2}) in 1-D
\begin{align}\label{state-2-3}
\left| \Psi \right\rangle = \prod_{s,i}\delta \left( \omega_{s}+\omega _{i} - \omega _{p}\right) 
 \, | \alpha_{s}(\omega_{s}) \rangle | \alpha_{i}(\omega_{i}) \rangle,
\end{align}
indicating a vector in the space of coherent state. 

Based on the entangled coherent states of Eq.~(\ref{state-2}), we now calculate its second-order coherence function 
$G^{(2)}(\mathbf{r}_1, t_1; \mathbf{r}_2, t_2)$ by assuming a simple measurement in which the signal beam and the idler beam, respectively, are directed to two point-like photodetectors $D_1$ and $D_2$ at distance. These detectors measure the intensities of the signal beam and the idler beam as functions of time, respectively, and are used to measure the intensity correlation between the signal beam and the idler beam, jointly.  It is no surprise that both $D_1$ and $D_2$ observe constant intensities, respectively, as expected from a CW laser beam. In the joint measurement of intensity correlation of two CW laser beams, one may ask if there is any surprise? 

In this 1-D measurement, it is reasonable to concentrate our calculation to the temporal correlation $G^{(2)}(r_1, t_1; r_2, t_2)$. 
We therefore simplify the state described by Eq.~(\ref{state-2}) to its 1-D version of Eq.~(\ref{state-2-3}).
The second-order temporal coherence or correlation function of the entangled bi-mode coherent state
is calculated as 
\begin{align}
& \ \ \ \ G^{(2)}(r_1, t_1; r_2, t_2) \cr
& = \langle \Psi | \,\hat{E}^{(-)}(r_{1}, t_{1}) 
\hat{E}^{(-)}(r_{2}, t_{2}) \hat{E}^{(+)}(r_{2}, t_{2}) \hat{E}^{(+)}(r_{1}, t_{1}) \, | \Psi\rangle \cr
& = |\, \langle \, \Psi \,|\, \hat{E}^{(+)}(r_{2}, t_{2}) \hat{E}^{(+)}(r_{1}, t_{1})\,|\, \Psi \, \rangle\,|^{2} \cr
& \equiv | \, \Psi(r_{1}, t_{1};  r_{2}, t_{2}) \, |^{2} 
\end{align}
where, again, we assume $D_1$ is triggered by one or more indistinguishable photons from the signal beam at $(r_1,t_1)$ while $D_2$ is triggered by one or more indistinguishable photons from the idler beam at $(r_2,t_2)$. $\hat{E}^{(-)}$ and $\hat{E}^{(+)}$ are the negative-frequency and the positive-frequency field operators at space-time points $(r_{1}, t_{1})$ and 
$(r_{2}, t_{2})$;
$\Psi(r_{1}, t_{1}; r_{2}, t_{2})$ is the probability amplitude to observe a joint photodetection event at space-time $(r_{1}, t_{1})$ and $(r_{2}, t_{2})$.  We name it the effective wavefunction of the entangled signal-idler laser beams \cite{ShihBook}.  
To evaluate $\Psi(r_{1}, t_{1}; r_{2}, t_{2})$
and $G^{(2)}(r_{1}, t_{1};  r_{2}, t_{2})$, we need to propagate the field operators from the source 
to space-time coordinates $(r_{1}, t_{1})$ and  $(r_{2}, t_{2})$.

In general, the field operator $\hat{E}^{(+)}(r, t)$ at space-time point $(r, t)$ can be written 
in terms of the Green's function which propagates each Fourier mode from
space-time point $(r_0, t_0)$ to $(r, t)$:
\begin{align}\label{Green-0}
\hat{E}^{(+)}(r, t) = \sum_{\omega} \, g({\omega}; r- r_0, t - t_0) \,
\hat{E}^{(+)}({\omega}, r_0, t_0).
\end{align}
To simplify the notation, we have assumed one polarization in Eq.~(\ref{Green-0}). 
Under this assumption, the field operators become
\begin{align}\label{gg-1}
\hat{E}^{(+)}(r_j, t_j) 
&\cong \sum_\omega \, 
e^{-i\omega \tau_{j}}\hat{a}(\omega) 
\end{align}
where $\tau_j \equiv t_j - r_j/c$ , 
$r_j$ is the longitudinal coordinate of the $jth$ photodetector.  We have chosen 
$r_0 =0$ at the output plane of the entangled bright light source.  For convenience, 
all constants associated with the field are absorbed into 
$g(\omega; r_j, t_j)$.  
Since $|\alpha(\omega)\rangle$ is an eigenstate of the annihilation operator with an eigenvalue $\alpha(\omega)$ \cite{Glauber}
$$
\hat{a}(\omega)|\alpha(\omega) \rangle=\alpha(\omega)|\alpha(\omega) \rangle, \
\hat{a}(\omega) |\Psi \rangle = \alpha(\omega) |\Psi \rangle, $$
the effective wavefunction $\Psi(\tau_1, \tau_2)$ of the entangled coherent state is calculated as 
\begin{align}
&\ \ \Psi(\tau_1, \tau_2) 
= \langle \, \Psi \,| \sum_{\omega, \omega'} 
e^{-i\omega \tau_{2}} \hat{a}(\omega) \, e^{-i\omega' \tau_{1}} \hat{a}(\omega') 
 |\, \Psi\, \rangle \\
 & = \Psi_0 \sum_{\omega_s, \omega_i} \delta(\omega_s + \omega_i - \omega_p) \,
\alpha(\omega_s) e^{-i\omega_s \tau_{1}} \, \alpha(\omega_i) e^{-i\omega_i \tau_{2}} \nonumber
\end{align}
indicating the coherent superposition of a large number of entangled amplitudes of the signal-idler pairs. 

The coherent superposition can be easily calculated with the help of the $\delta$-function,
\begin{align}
& \ \ \Psi(\tau_1, \tau_2) 
=  \Psi_0 \sum_{\omega_s, \omega_i} \delta(\omega_s + \omega_i - \omega_p) \,
\alpha(\omega_s) \alpha(\omega_i) \cr
& \hspace{15mm} \times e^{-i \frac{1}{2}(\omega_s + \omega_i)(\tau_{1}+ \tau_2)} \, 
e^{-i \frac{1}{2}(\omega_s - \omega_i)(\tau_1- \tau_{2})} \cr
& =  \Psi_0 \, e^{-i \omega_p(\tau_{1}+ \tau_2)/2} \, \sum_{\nu}  \,
f(\nu ) \,  e^{-i \nu(\tau_1- \tau_{2})} 
\end{align}
where we have assumed degenerate OPO, i.e., $\omega^0_{s} = \omega^0_{i}$ with $\omega^0_{s}$ and $\omega^0_{i}$ 
the central frequencies of the signal and the idler, $\nu$ is the detuning frequency $\nu = \omega_s - \omega^0_{s}$. 

Furthermore, we can assume an OPO with a large number N of cavity-modes, 
$\nu \equiv n \omega_b$, $n = 0, 1, ..., N-1$, and
a constant mode distribution function $f(\nu)$,
\begin{align}\label{wavepacket-1}
& \ \ \Psi(\tau_1, \tau_2)
\simeq \Psi_0 \, e^{-i \omega_p(\tau_{1}+ \tau_2)/2} \, \sum_{n=0}^{N-1}  \, e^{-i n\omega_b (\tau_1- \tau_{2})} \cr
&= \Psi_0 \, e^{-i \omega_p(\tau_{1} + \tau_{2})/2} \,
\big{[} e^{i (N-1) \omega_b \tau /2} \,
\frac{\textrm{sin} \, N \omega_b \tau /2}{\textrm{sin} \, \omega_b \tau/2} \big] 
\end{align}
where $\omega_b$ is the beating frequency between neighboring cavity-modes, 
$\tau \equiv \tau_1 - \tau_2$.  
In the above calculation, we have assumed each mode contains a single-frequency. Realistically, 
we may have to take into account the finite spectral bandwidth of each mode \cite{ShihBook},
\begin{align}
& \ \ \Psi(\tau_1, \tau_2) 
= \langle \Psi | \sum_{\omega', \omega"} 
e^{-i\omega' \tau_{2}} \hat{a}(\omega') e^{-i\omega" \tau_{1}} \hat{a}(\omega") 
 |\Psi\rangle \cr
 & = \Psi_0 \sum_{\omega_s, \omega_i} \delta(\omega_s + \omega_i - \omega_p) \,
{\mathcal F}_{\tau_1} \big\{ \alpha(\omega_s)\big\} e^{-i\omega_s \tau_{1}} \cr
& \ \ \ \ \times {\mathcal F}_{\tau_2} \big\{ \alpha(\omega_i)\big\} e^{-i\omega_i \tau_{2}}  \cr
& =  \Psi_0 \sum_{\omega_s, \omega_i} \delta(\omega_s + \omega_i - \omega_p) \,
{\mathcal F}_{\tau_1} \big\{ \alpha(\omega_s)\big\} {\mathcal F}_{\tau_2} \big\{ \alpha(\omega_i)\big\} \cr
& \hspace{15mm}\times e^{-i \frac{1}{2}(\omega_s + \omega_i)(\tau_{1}+ \tau_2)} \, 
e^{-i \frac{1}{2}(\omega_s - \omega_i)(\tau_1- \tau_{2})} \cr
&= \Psi_0 e^{-i \omega_p(\tau_{1} + \tau_{2})/2} 
\sum_{n=0}^{N-1} 
{\mathcal F}_{\tau_1} \big\{ \alpha(\omega^0_s + n \omega_b) \big\} \cr
& \times {\mathcal F}_{\tau_2} \big\{ \alpha(\omega^0_i -n \omega_b)\big\} 
\Big{[} e^{i (N-1) \omega_b \tau /2} \,
\frac{\textrm{sin} \, N \omega_b \tau /2}{\textrm{sin} \, \omega_b \tau/2} \Big] 
\end{align}
where ${\mathcal F}_{\tau_1} \big\{ \alpha(\omega^0_s + n \omega_b)\big\} $ and 
${\mathcal F}_{\tau_2} \big\{ \alpha(\omega^0_i -n \omega_b)\big\}$ are the Fourier transforms of the spectral functions of the 
signal and idler cavity-modes, respectively. It is clear, the amplitude of the effective wavefunction is determined by two functions: (1) the value of $\textrm{sin} Nx / \textrm{sin} x$; and
(2) the degree of overlap between the Fourier transforms ${\mathcal F}_{\tau_1} \big\{ \alpha(\omega^0_s + n \omega_b)\big\} $ and 
${\mathcal F}_{\tau_2} \big\{ \alpha(\omega^0_i -n \omega_b)\big\}$. For a large number of cavity-modes, the summation can be replaced by a convolution integral 
\begin{align}
& \ \ \ \sum_{n=0}^{N-1}  \,
{\mathcal F}_{\tau_1} \big\{ \alpha(\omega^0_s + n \omega_b)\big\} {\mathcal F}_{\tau_2} \big\{ \alpha(\omega^0_i -n \omega_b)\big\} \cr
& \simeq \int d t_{0n} \, {\mathcal F}_{\tau_{1n}} \big\{ \alpha(\omega_s)\big\} {\mathcal F}_{\tau_{2n}} \big\{ \alpha(\omega_i)\big\} \cr
& \simeq {\mathcal F}_{\tau} \big\{ \alpha^2(\omega) \big\}
\end{align}
where $\alpha(\omega)$ specifies the spectral distribution of the cavity-modes, $t_{0n}$ is the creation time of the $n$th OPO 
cavity-mode for which we have assumed a random distribution of $t_{0n}$. We thus have
\begin{align}
G^{(2)}(\tau) 
\propto \big| {\mathcal F}_{\tau} \big\{ \alpha^2(\omega) \big\} \big|^2 \Big[ \frac{\textrm{sin}^2 \, 
N \omega_b \tau /2}{\textrm{sin}^2 \, \omega_b \tau/2} \Big] .
\end{align}
Assuming a constant distribution function $a(\omega)$ for each cavity-mode within its spectral bandwidth $\Delta \omega$,
and for all cavity-modes, 
\begin{align}
G^{(2)}(\tau) 
\propto \Big[ \textrm{sinc}^2 \frac{\Delta \omega \tau}{2} \Big]\, \Big[  
\frac{\textrm{sin}^2 \, N \omega_b \tau /2}{\textrm{sin}^2 \, \omega_b \tau/2} \Big]
\end{align}
where $\Delta \omega$ is the spectral bandwidth, or line-width, of the cavity-mode, again, 
$\tau \equiv \tau_1 - \tau_2 = (t_1 - t_2) - (r_1- r_2)/c$.  

\begin{figure}
\centering
\includegraphics[width = 1.05\linewidth]{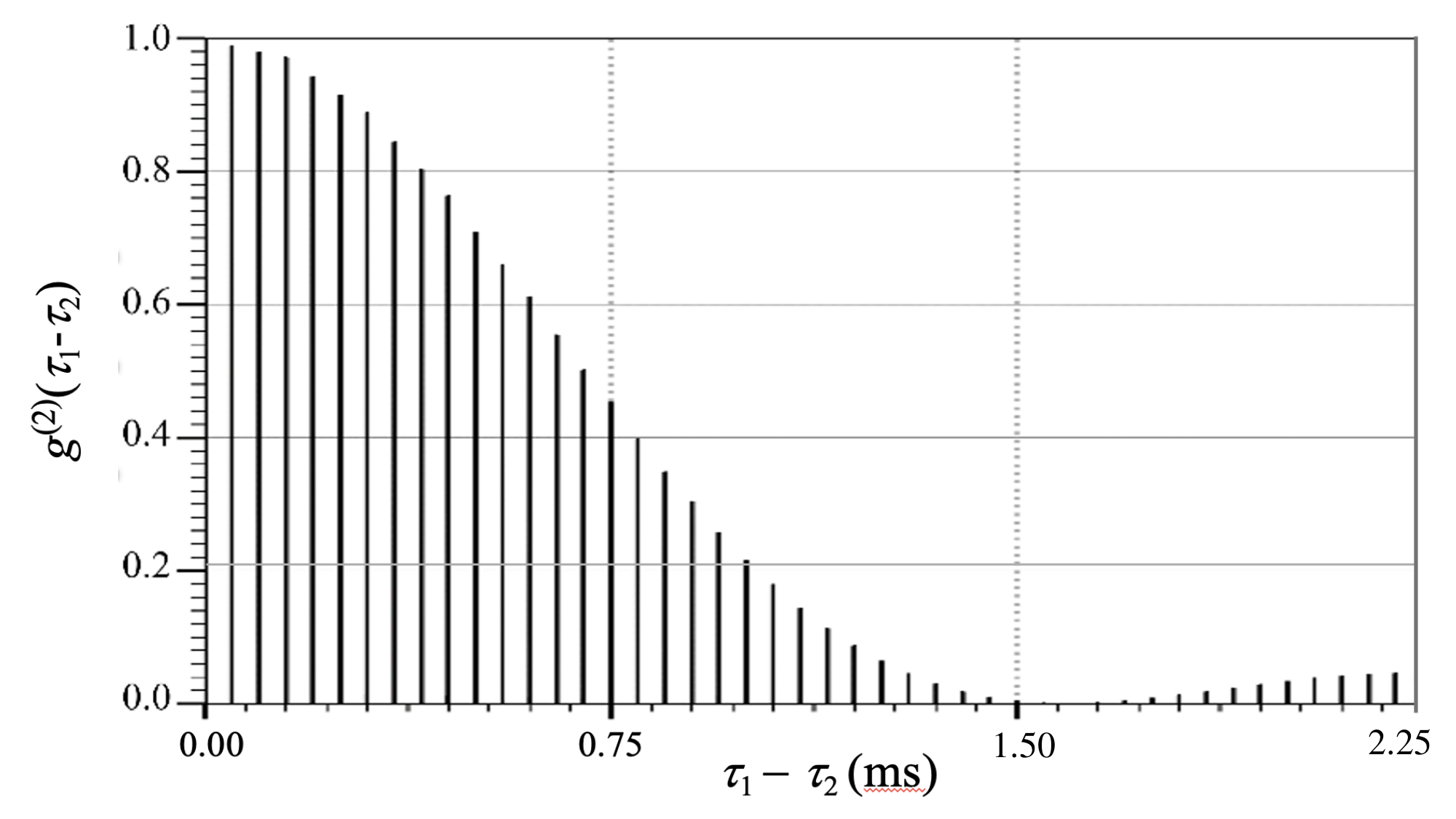}
\caption{Numerical simulation of a 100\% contrast QGFC:  comb-function with sinc-function envelope observable in the joint photodetection of the entangled signal-idler beams, although both $D_1$ and $D_2$ measure constant intensities as expected from CW laser beams. In this simulation, we assumed 100,000 cavity modes with mode separation $\omega_b /2 \pi = 20$ kHz. We also assumed finite spectral bandwidth of 200 Hz for each cavity mode which results in a sinc-function envelope of 1.50 ms temporal width on top of the comb-function. The temporal width of each comb peak is 500 ps. } 
\label{fig:GFC}
\end{figure}
The QGFC has multipole periodic sharp correlation peaks at $\omega_b \tau /2 =n \pi$, for $n = 0, \pm 1, \pm 2, ..., \pm (N-1)$, 
\begin{align}\label{Fitting}
(t_1 - t_2)_n =  \frac{1}{\nu_b} \, n + (r_1 - r_2)/c
\end{align}
where we have used $\omega_b = 2\pi \nu_b$ and defined $(t_1 - t_2)_n$ as the measured value of $t_1 - t_2$ at the $n$th comb peak. The temporal width of each QGFC peak, measured between neighboring zeros of the correlation function, is approximately 
\begin{align}
\Delta t \simeq \frac{1}{\nu_b N}.
\end{align}
Figure~\ref{fig:GFC} illustrates a numerical simulation of a typical QGFC observable from the nonlocal correlation measurement of entangled laser beams.  

Indeed it is surprising that a 100\% contrast comb-like function is observable from the joint-detection of the entangled signal-idler laser beams, although both detectors, $D_1$ and $D_2$, measure constant light intensities.  We are facing two serious questions: (1) What is the cause of these periodic sharp correlation?  (2) Can we trust zero-coincidences, or anti-correlation, in the joint measurement of CW laser beams?    

We know that most of the observed correlation functions of entangled photon pairs, either created from atomic cascade decay or generated from SPDC, are Gaussian-like functions with a single peak at $\tau_1 - \tau_2 =0$.  A widely accepted idea is that this is due to the simultaneous generation of entangled photon pairs at the source. Since the entangled photons are produced simultaneously in the source, they must be jointly detected by the detectors at $\tau_1 - \tau_2 =0$. Following 
this line of thought, can we assume that the entangled signal-idler laser beams are 
generated simultaneously and periodically by the CW OPO? Not likely! CW OPOs do not work that way. 

The contrast of the QGFC is 100\%, which means the signal-idler laser beams are anti-correlated, i.e., no joint-detection event occurs whenever the relative delay of $\tau_1 - \tau_2$ falls into the region between the periodic sharp correlation peaks. From the perspective of classical thinking, it is absolutely impossible for zero-coincidence to occur in the measurement of CW laser beams. Even if somehow there are chances to observe zero-coincidences, why are they periodic in such a peculiar manner? 

The QGFC is the result of a standard and straightforward calculation based on the quantum theory of nonlocal superposition, similar to what Einstein-Podolsky-Rosen did in 1935 \cite{EPR}.  The periodic QGFC looks like a multi-slit diffraction pattern. The comb-like multi-slit diffraction pattern is the result of coherent superposition of a large number of radiation fields coming from the narrow multi-slits. The ``zeros" and periodic sharp peaks of the multi-slit diffraction pattern are the results of destructive and constructive interference.  Similarly, the QGFC is the result of coherent superposition of a large number of quantum amplitudes, corresponding to the nonlocal joint-measurement of a large number of entangled signal-idler pairs. The ``zeros" and periodic peaks of the QGFC are also the results of destructive and constructive interference, except that this interference occurs at distant space-time coordinates $(z_1, t_1)$ and $(z_2, t_2)$, and the two photodetection events can be easily managed to achieve space-like separation.  Apparently, the observation of QGFC is strong evidence of quantum nonlocal superposition, or nonlocal interference. 

It is worth noting that the above discussions are based on idealized conditions. The contrast of experimentally observed QGFC may drop below 100\% due to imperfect phase matching and partially coherent states caused by photon losses. 

In summary, this letter reports on the study of entangled laser beams, or entangled coherent states, from their generation to nonlocal correlation-anticorrelation. The entangled coherent states can be produced by pumping a multi-longitudinal-cavity-mode OPO with a CW monochromatic laser beam, operating at a level well above the threshold. The peculiar behavior of QGFC verifies the quantum nonlocal superposition, which is fundamentally interesting and practically useful. Superior to entangled photon pairs, measurements of entangled laser beams do not rely on photon counting and can be performed over greater distance in shorter time with higher resolution and accuracy.

The author would like to thank T. A. Smith, M. F. Locke, B. Joshi, F. A. Narducci and M. Fitelson for helpful discussions.

\end{document}